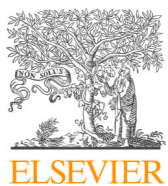
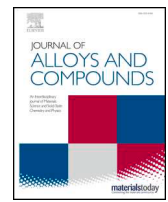

Review

# CoSb$_3$-based skutterudite nanocomposites prepared by cold sintering process with enhanced thermoelectric properties

Aida Serrano [a],[*],[1], Olga Caballero-Calero [b], Cecilia Granados-Miralles [a], Giulio Gorni [c], Cristina V. Manzano [b], Marta Rull-Bravo [b], Alberto Moure [a], Marisol Martín-González [b], José F. Fernández [a]

[a] *Departamento de Electrocerámica, Instituto de Cerámica y Vidrio (ICV), CSIC, E-28049 Madrid, Spain*
[b] *Instituto de Micro y Nanotecnología, IMN-CNM, CSIC (CEI UAM+CSIC), Isaac Newton 8, E-28760, Tres Cantos, Madrid, Spain*
[c] *CELLS-ALBA Synchrotron Light Facility, E-08290 Cerdanyola del Vallès, Spain*

## ARTICLE INFO

## ABSTRACT

We show here for the first time the use of a cold sintering process (CSP) to sinter CoSb$_3$-based thermoelectric materials. CSP at 150 °C for 90 min under a uniaxial pressure of 750 MPa yields pieces with a relative density of 86 %, which is increased to around 92 % after a post-annealing at temperatures ≥ 500 °C in Ar atmosphere. The reported CSP produces Te doped-CoSb$_3$ nanocomposites with similar morphological and structural characteristics to the starting nanopowders obtained by ball milling in air atmosphere. The post-thermal treatment induces grain coalescence and grain growth, crystallite size growth as well as compositional changes in the nanocomposite, decreasing the amount of the main phase, CoSb$_3$, and increasing the weight of secondary phase, CoSb$_2$, up to a 30 wt% at 600 °C. Remarkably, the average valence for the Co, Sb and Te absorbing atoms is neither transformed by the sintering process nor by the subsequent heat treatment. The functional response of the sintered thermoelectric nanocomposites exhibits a maximum figure of merit of 0.12(3) at room temperature for the nanocomposites sintered by CSP with a subsequent post-annealing at 500 °C. This is mainly due to its low thermal conductivity in comparison with similar powders sintered by other approaches, and it is explained by the morphological and structural properties. These findings represent an attractive alternative for obtaining efficient thermoelectric skutterudites by a scalable and cost-effective route.

## Contents

[*] Corresponding author.
  *E-mail address:* aida.serrano@icv.csic.es (A. Serrano).
[1] ORCID: 0000-0002-6162-0014.

## 1. Introduction

The skutterudites form a large class of compounds with many novel properties depending on their composition [1,2]. Particularly, skutterudites usually exhibit potential thermoelectric properties due to a high Seebeck coefficient and a reasonable electrical conductivity, but a relatively high thermal conductivity [3–7]. Many of the interesting properties of skutterudites as thermoelectric materials depend on their unusual cubic crystal structure, containing two large voids often filled with a variety of atoms: e.g., rare earths, alkaline earths, etc. These can act as rattlers and thus, reduce the thermal conductivity of the structure, being a great example of the so-called electron crystal phonon glass materials [8–13].

The general formula of skutterudites is $TPn_3$ where T is a transition metal and Pn is a pnictogen. The skutterudite compounds that have been most intensively studied are the antimonides. Specifically, $CoSb_3$-based materials show a skutterudite-type structure with a great potential for thermoelectric applications [14–16]. $CoSb_3$ holds a high electrical conductivity and a high Seebeck coefficient, but its thermal conductivity is also high, which limits its thermoelectric performance. Therefore, many efforts are being focused on reducing this parameter to obtain high values of the figure of merit ($zT$) [14,17], such as multi-filling [18], nanostructuring or gain boundary engineering, etc [19]".

Usually, the efficiency of $CoSb_3$ can be improved following several strategies, such as by modifying the crystalline structure, doping, or by control of composition or structure at the nanoscale, where defects act as grain boundaries that can help to enhance phonon scattering and reduce thermal conductivity [17]. For these proposes, a first step usually comprises the use of cost effective and/or easy methods [20] to have nanopowders with the adequate composition. To retain the nanostructuration, several syntheses and sintering methods have been investigated. For example, hot pressing (HP), high-pressure torsion (HPT) or sintering assisted by spark plasma sintering (SPS) have been proven as densification techniques that, to a certain extent, maintain the nanostructuration in the bulk material. As an example, in a previous work we reported Te doped-$CoSb_3$ nanocomposites prepared by high-energy ball-milling mechanosynthesis and sintered by SPS, obtaining a nanostructured bulk with an improved thermoelectric response related to interface effects [6]. However, these processing methods are expensive and they present unsolved problems associated with later industrial production. An easily scalable sintering route consisting of a simple pressing and annealing sintering process was followed for the same Te doped-$CoSb_3$ mechanosynthesized nanopowders as in reference [6]. This method looked like a promising cost-effective route [21], although the performance did not reach yet that of SPS sintered nanocomposites.

In this context, the recently developed cold sintering process (CSP) may be considered an alternative route to obtain nanostructured $CoSb_3$-based composites. CSP consists of the sintering of a compound using at least one inorganic compound in powder form which is mixed with a solvent that partially solubilizes the particles to promote their mass transport, and then applying a uniaxial pressure at low temperatures (< 250 °C), to make it into a dense material [22–27]. Nowadays, CSP has been employed to sinter a large number of compounds, from piezo-electric materials up to ceramic magnets [23,28–30], reducing temperatures and costs in the sintering processes. Certain thermoelectric materials have been prepared following CSP, such as $Ca_3Co_4O_9$ ceramics [31,32] and $Bi_2Te_3$-based compounds [33], and the results and the functional properties obtained through this route are quite promising. However, no studies on CSP of skutterudite-based materials have not yet been conducted, in part due to the limitation of using a water-based liquid medium in metal-based nanopowders.

Here, we have sintered skutterudites based on $CoSb_3$ by a route assisted by CSP using glacial acetic acid as organic solvent plus a subsequent post-annealing process under Ar atmosphere. For such purpose, mechanosynthesized Te doped-$CoSb_3$ nanopowders as in reference [6] were employed. Morphological, structural and thermoelectric properties at room temperature (RT) of the resulting nanocomposites are studied. Controllable relative density, composition, grain size, crystalline size and short-range order are analyzed and related to the physical properties. The thermoelectric response reaches the highest $zT$ values for the nanocomposite prepared by CSP plus a subsequent post-annealing at 500 °C close to the operating temperature, according to the compositional and structural characteristics of the nanocomposites and outperforming the previously reported values.

## 2. Materials and methods

$CoSb_3$-based nanopowders were prepared following the procedure described in a previous work [6]. Briefly, a mixture of stoichiometric amounts to obtain $CoSb_3$ composition, from Sb (purity of 99.8 %) in powder form and Co precursor consisting of a mixture of 50 wt% metallic Co (purity of 99.5 %) and 50 wt% $Co_3O_4$, was doped with 15 wt% Te and milled for 14 h in air atmosphere in a tungsten carbide (WC) pot with five WC balls (2 cm diameter, ball-to-powder weight ratio ~ 30/1). The powder mixture was mechanosynthesized with a planetary mill (Fritsch Pulverisette 6) operating at 300 rpm. Then, the final powders were processed by CSP in a Burkle D-7290 press using glacial acetic acid as solvent. The amount of solvent added was 50 wt% of the total mass of powders. For that, powders were manually mixed for 10 min in an agate mortar with the glacial acid acetic to obtain a granulated powder, which was pressed at 220 MPa for 5 min at RT in a cylindrical die with an inner diameter of 0.83 cm. Subsequently, the pressed powders were sintered to CSP heating at a target temperature of 150 °C for 90 min and applying a pressure of 750 MPa. The annealing rate of 20 °C/min was used, and after dwell time at the set temperature, the CSP pellets were let to naturally cool down in air. Prior to the cold sintering method, processing parameters such as the height/diameter of pellets, target temperature, pressure, and solvent wt% were optimized to obtain the highest density values of the nanocomposites (not shown here). Finally, a post-annealing in Ar atmosphere was followed for 120 min varying the target temperatures at 300, 400, 500 and 600 °C with 5 °C/min rates for both heating up and cooling down. Table 1 shows the samples selected for this work indicating their sintering conditions.

The density values of the sintered nanocomposites were determined by the Archimedes method and by the mass/dimension

**Table 1**

Description of nanocomposites sintered for this work, indicating the sintering conditions and the calculated relative density, considering the phase composition extracted from Rietveld analysis of PXRD data. CSP stage is achieved at 150 °C for 90 min. The corresponding post-annealing process is carried out at target temperatures of 300, 400, 500 or 600 °C for 120 min in Ar atmosphere.

| Sample | Sintering conditions | Density (g cm$^{-3}$) | Relative density (%) |
| --- | --- | --- | --- |
| **NC** | CSP | 6.25(3) | 86(1) |
| **NC 300** | CSP + 300 °C | 6.45(5) | 89(1) |
| **NC 400** | CSP + 400 °C | 6.58(3) | 91(1) |
| **NC 500** | CSP + 500 °C | 6.70(3) | 92(1) |
| **NC 600** | CSP + 600 °C | 7.21(3) | 92(1) |

values. The relative density was calculated considering theoretical density values of 7.59 g cm$^{-3}$ for $CoSb_3$, 8.29 g cm$^{-3}$ for $CoSb_2$, 6.70 g cm$^{-3}$ for Sb, 5.84 g cm$^{-3}$ for $Sb_2O_3$, 5.70 g cm$^{-3}$ for $CoSb_2O_4$ and 8.90 g cm$^{-3}$ for Co [34–37], and the relative weight of each compound was estimated from Rietveld analysis of powder X-ray diffraction (PXRD) data.

The grain morphology of the starting powder and nanocomposites was studied by field emission scanning electron microscopy (FESEM) with an S-4700 Hitachi instrument at 20 kV. Sintered pieces were analyzed on freshly fractured surfaces. ImageJ software [38] was used to analyze the average particle size from the particle size distribution obtained from the FESEM images.

The electrical conductivity ($\sigma$) at RT was measured in-plane with a commercial Hall HMS-5500 system (from Ecopia, with an associated error of 5 %). The Seebeck coefficient ($S$) was also measured in-plane configuration with a lab-developed system, which allows the measurement of a temperature gradient around RT and the voltage produced. The experimental error associated with this measurement is 10%. Finally, the thermal conductivity ($k$) measurements were collected in the out-of-plane direction using a custom-made photoacoustic system [39,40]. In this technique, the acoustic waves produced by the periodic heating of the surface of the sample (produced by a modulated laser of 980 nm of wavelength) are detected by a microphone. The measurement for bulk samples, as in this work, is made in two different modes: heating and detecting on the same surface of the sample (reflection configuration) and heating on one side of the sample and detecting the acoustic waves on the other one (transmission configuration). From the phase shift between the laser pulse and the detected acoustic signal and the comparison between rear and front configuration, the thermal diffusivity, $\alpha$, is obtained. The thermal conductivity is determined from the relation $k = \alpha \rho c_p$, being $\rho$ the density and $c_p$ the specific heat. More details can be found in reference [21]. It is worth noting that for the measurement in transmission configuration, the samples must be thin enough to transmit the heat from one side to the other. To that end, the samples were polished until thicknesses of around 300 μm were achieved. Taking all of these into account, the associated error of the thermal conductivity measurements will be taken as 15 %.

The structural characterization of samples was carried out by PXRD and X-ray absorption spectroscopy (XAS) experiments. The crystalline phases of the initial products and sintered samples were characterized by PXRD (D8, Bruker) using a Lynx Eye detector and a Cu K$\alpha$ radiation with a $\lambda = 0.154$ nm in the 2$\theta$ range of 10–80°. Rietveld refinements of the PXRD data were carried out using the software FullProf [41]. In the refinements, a Thompson-Cox-Hastings pseudo-Voigt function was used to model the peak-profile [42]. The instrumental contribution to the peak-broadening was determined from measurements of a standard powder (NIST LaB6 SRM® 660b) [43] and deconvoluted from the data. The sample contribution to the broadening was considered as purely size-originated. In all nanocomposites, quantitative information on the crystalline compositional phases, the lattice parameters and the average crystallite size was extracted. The uncertainty in the refined parameters is calculated considering the propagation of errors. The phases included in the models were $CoSb_3$ (cubic, $Im$-3), $CoSb_2$ (monoclinic, $P2_{1/c}$), $CoSb_2O_4$ (tetragonal, $P4_2/mbc$), $Sb_2O_3$ (orthorhombic, $Pccn$), Sb (trigonal, $R$-3$m$) and Co (cubic, $Fm$-3$m$) [34–37].

XAS measurements were performed on the CSP and post-annealed composites to investigate the effect of the sintering process on the oxidation state and short-range order of the structures. Measurements were carried out at the BL22 CLÆSS beamline of the ALBA synchrotron facility in Cerdanyola del Vallès (Spain) [44]. Both X-ray absorption near-edge structure (XANES) and extended X-ray absorption fine structure (EXAFS) experiments were achieved at RT and in transmission mode, at the Co (7709 eV), Sb (30491 eV) and Te (31814 eV) K-edges. The powdered samples were mixed with cellulose and pressed into pellets. The monochromator used in the experiments was a double Si crystal oriented in the (311) direction and the beam size at the sample position was around 400 (H) × 300 (V) μm$^2$. The incident and transmitted intensities were detected by two ionization chambers. Co and Sb metal foils were measured in transmission configuration and used to calibrate the energy. Starting $CoSb_3$-based powders before the sintering process were measured as a reference as well as $Co_3O_4$, CoO, Co, $Sb_2O_3$, $Sb_2O_5$, Sb, $TeO_2$ and Te references from transmitted photons.

## 3. Results and discussion

### 3.1. Density and morphological properties

CSP is a sintering route in which a large number of parameters can be evaluated and optimized [22,26]. This mouldable technique is advantageous both for densifying materials in a low cost and controlled way as well as for tailoring the final properties of the sintered pieces. Here, after the optimization process of CSP parameters (not shown), nanocomposites based on $CoSb_3$ with a density value of 6.25 g cm$^{-3}$ are obtained from the starting powders got by the mechanosynthesis process and submitted to CSP. Considering the phase composition calculated by PXRD, a relative density of 86 % is achieved. Thus, well-consolidated nanocomposites are obtained after the sintering process at 150 °C and without reaching high temperatures such as those commonly used in conventional sintering processes. The density of the CSP pieces can be further increased by a post-annealing process in Ar atmosphere, with values of 6.45 g cm$^{-3}$ at 300 °C, 6.58 g cm$^{-3}$ at 400 °C, 6.70 g cm$^{-3}$ at 500 °C and 7.21 g cm$^{-3}$ at 600 °C of post-thermal treatment. Table 1 shows the density and the relative density values obtained for the different sintered samples. It should be noted that the pieces post-annealed at 500 °C and 600 °C show the same relative density with respect to the corresponding theoretical values, in spite of presenting different density values. These findings are due to the compositional changes related to the post-annealing temperature, as shown below.

The morphological characteristics of the thermoelectric pieces sintered by CSP, after the post-annealing process as well as the starting powders are displayed in Fig. 1. The FESEM micrograph of starting powders in Fig. 1a shows nanoparticles with rounded morphology and an average particle size of 24(3) nm. In addition, agglomerates of nanoparticles are identified with a size ranging from 100 to 1000 nm, similar to our previous work [6]. Fig. 1b–f show the fresh fracture of composites sintered by CSP along with those prepared by CSP plus a post-annealing process at different temperatures. For the nanocomposite sintered by CSP (NC), similar grain size is identified with respect to the starting powder of around 26(2) nm preserving the rounded shape (see Fig. 1b), indicating no change in particle size during the CSP. For the sample post-annealed at 300 °C (NC 300) the grain growth is limited, with an average size of 43(6) nm (Fig. 1c). However, two main features should be noted: a homogeneous grain distribution and the appearance of grain boundaries. For NC 400, an increase in the average size is observed with a value of around 53(11) nm (Fig. 1d). As the post-annealing process is followed at 500 °C (see Fig. 1e), a clear growth of the grain size is obtained with irregular grains having faceted grain boundaries. The average grain size is around 195(34) nm and a low presence of porosity is identified. Finally, for sample NC 600 (post-annealed at 600 °C), an exaggerated grain growth is obtained, with an average size of around 733(119) nm (see Fig. 1f). In the last two samples, the coalescence of smaller grains into large equiaxed grains is identified and the presence of triple point junctions having 120° angles and straight grain boundaries indicates that the microstructural equilibrium is reached. The average grain sizes for all nanocomposites are displayed in Fig. 1g, where an exponential growth of grain size with the post-annealing temperature is revealed. These results can be correlated with the measured density values, indicating that higher densification takes place with the greater grain coarsening occurring at the higher post-annealing temperatures.

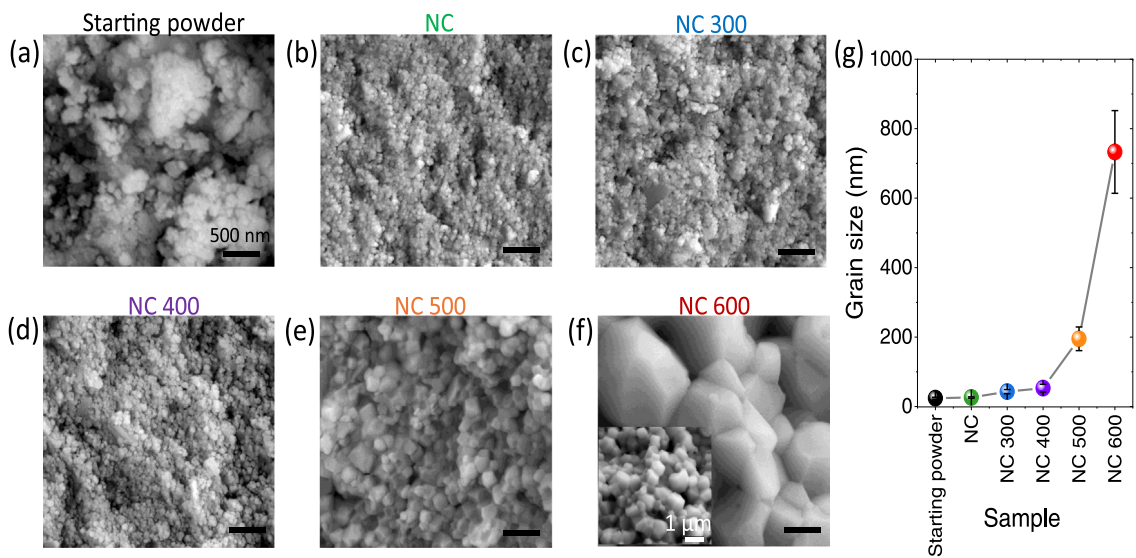

Fig. 1. FESEM micrographs of (a) the starting powder, and fresh fracture surface of cold sintered ceramics (b) NC, (c) NC 300, (d) NC 400, (e) NC 500 and (f) NC 600. A micrographic for sample NC 600 with a lower magnification is also shown on panel (f). (g) Average grain size for the starting powders along with NC, NC 300, NC 400, NC 500 and NC 600, showing an exponential grain growth.

### 3.2. Thermoelectric response

To study the thermoelectric properties of the different $CoSb_3$-based nanostructured bulk materials sintered by CSP and evaluate the influence of the thermal annealing, the electrical conductivity ($\sigma$), Seebeck coefficient ($S$) and thermal conductivity ($k$) were measured at RT. The obtained values for the different parameters as a function of the annealing temperature of the post CSP treatment are shown in Fig. 2 and compared with the values published for the nanostructured-bulk materials sintered by SPS from the same starting powders [6].

The electrical conductivity, shown in Fig. 2a, has a minimum value of 500(30) S m$^{-1}$ for the CSP nanocomposite without any subsequent annealing. This value increases with the increasing annealing temperature, reaching the highest value over 90 times bigger than the starting one, of 46000(2000) S m$^{-1}$ for the nanocomposite sintered by CSP and post-annealed at 600 °C. When compared to the samples sintered by SPS, this value is half of that obtained in reference [6], which was 78700 S m$^{-1}$.

In the case of the Seebeck coefficient (see Fig. 2b), the nanocomposite NC has a value of − 100(10) μV K$^{-1}$, which increases to − 125(10) μV K$^{-1}$, and − 130(10) μV K$^{-1}$, for the nanocomposites post-annealed at 300 and 400 °C, respectively. Then, for the nanocomposite post-annealed at 500 °C, a maximum value of the Seebeck coefficient of − 200(20) μV K$^{-1}$ is found. Finally, for the nanocomposite NC 600 at the highest post-annealing temperature, the Seebeck coefficient decreases abruptly to a value of − 80(10) μV K$^{-1}$, even lower than for the untreated nanocomposite (sample NC). The reason behind this decrease is related with the content of $CoSb_3$ when compared to other phases in the samples, as it will be shown when describing the structural characterization. When compared to the SPS nanocomposites reported in Ref. [6], the maximum value of − 200 μV K$^{-1}$ obtained for the nanocomposite post-annealed at 500 °C is almost double (− 135 μV K$^{-1}$).

Then, when the power factor is calculated (Fig. 2c), it can be seen that the highest value corresponds to the sample NC 500, with a value of 1000(200) μW m$^{-1}$ K$^{-2}$. Comparing the values obtained for these CSP nanocomposites with those reported for the nanocomposite sintered by SPS from the same nanopowders in reference [6] (shown as dashed lines in Fig. 2a–c) the values are lower, with the maximum power factor obtained for the CSP nanocomposites of almost 40 % lower.

The measurement of the thermal conductivity obtained by the photoacoustic method is shown in Fig. 2d. In the case of the samples sintered by CSP (i.e. without post-thermal treatment) or with the lowest post-annealing treatment (300 °C, sample NC 300), the thermal conductivity could not be measured due to their low mechanical integrity that impeded reducing their thickness by polishing despite their density. From the remaining nanocomposites, it can be seen that the thermal conductivity increases with the temperature of the post-annealing treatment, with values ranging from 1.5(2) W m$^{-1}$ K$^{-1}$ to 3.6(5) W m$^{-1}$ K$^{-1}$. That trend is also in good agreement with the trend found for electrical conductivity.

Then, taking into account that skutterudites are isotropic systems, we can estimate the figure of merit of these samples, which corresponds to $zT = \sigma S^2 T \kappa^{-1}$, as shown in Fig. 2e, obtaining values at RT of 0.05(1) for the nanocomposite post-annealed at 400 °C and a maximum value of 0.12(3) for the nanocomposite post-annealed at 500 °C. The lowest value of 0.024(6) is obtained for the nanocomposite post-annealed at 600 °C. It is worth noting that the $zT$ value achieved for NC 500 is higher than that obtained for SPS sintered nanocomposites, which was reported to be 0.096 at RT in Ref. [6]. Also, this value is markedly higher than that obtained when the sintering process of the nanopowders was pressed and followed by thermal annealing, where values of $zT$ of 0.06 at RT were reported [21].

### 3.3. Structural characterization

Long-range structural characterization of the CSP sintered nanocomposites and after post-thermal treatment has been carried out by means of PXRD. Fig. 3 shows the PXRD patterns measured for the sample NC (prepared by CSP) and sample NC 400 (post-annealing at 400 °C), while the PXRD patterns of the rest of the samples are shown in the Supporting information (SI). Rietveld models to the PXRD data have been built without considering the Te doping, given that neighbouring elements such as Sb and Te are hardly distinguished from one another with X-ray diffraction [45]. In all cases, the major phase is $CoSb_3$, identifying a percentage of secondary phases that depends on the sintering process and post-treatment temperature. Refined weight fractions for each phase, retrieved from Rietveld refinements of the PXRD data, are represented in Fig. 4a. Exact weight fractions from each crystalline phase and sample may be found in Table S1 in SI. A similar amount of $CoSb_3$ (around 73 %) is determined after the CSP process in NC than in the starting powder, identifying the loss of Sb phase and the appearance of $CoSb_2O_4$ with a 3.4(4) % (see Table S1 in SI). Therefore, the CSP does not significantly modify the composition of the starting powder, which could have been a consequence of the use of glacial acetic acid as solvent and applying pressure during the sintering process, as has already been observed for other types of materials [28,30]. As the post-annealing process is followed in Ar atmosphere, similar compositions are obtained up to 500 °C (NC 500) for which a slight increase of the $CoSb_3$ (around 79 %) and $CoSb_2O_4$ (around 9 %) at the expense of $Sb_2O_3$ (7 %), as well as a decrease of $CoSb_2$ (around 5 %) are determined. The most important variations in the composition are achieved as the post-thermal temperature of the CSP sample is increased to 600 °C (NC 600). A reduction of the $CoSb_3$ to 67 %, an increase of the $CoSb_2$ to 30 % and the presence of some metallic Co as a minority phase (3 %) are obtained, which affect the relative density of the pieces as mentioned above. This composition transformation at 600 °C could be associated with a solid-gas driven reduction reaction inducing the decomposition of oxide into metallic phases based on the Ellingham diagram [46,47] or/and the sublimation of Sb increasing the amount of the $CoSb_2$ phase and decreasing that of $CoSb_3$ [48,49].

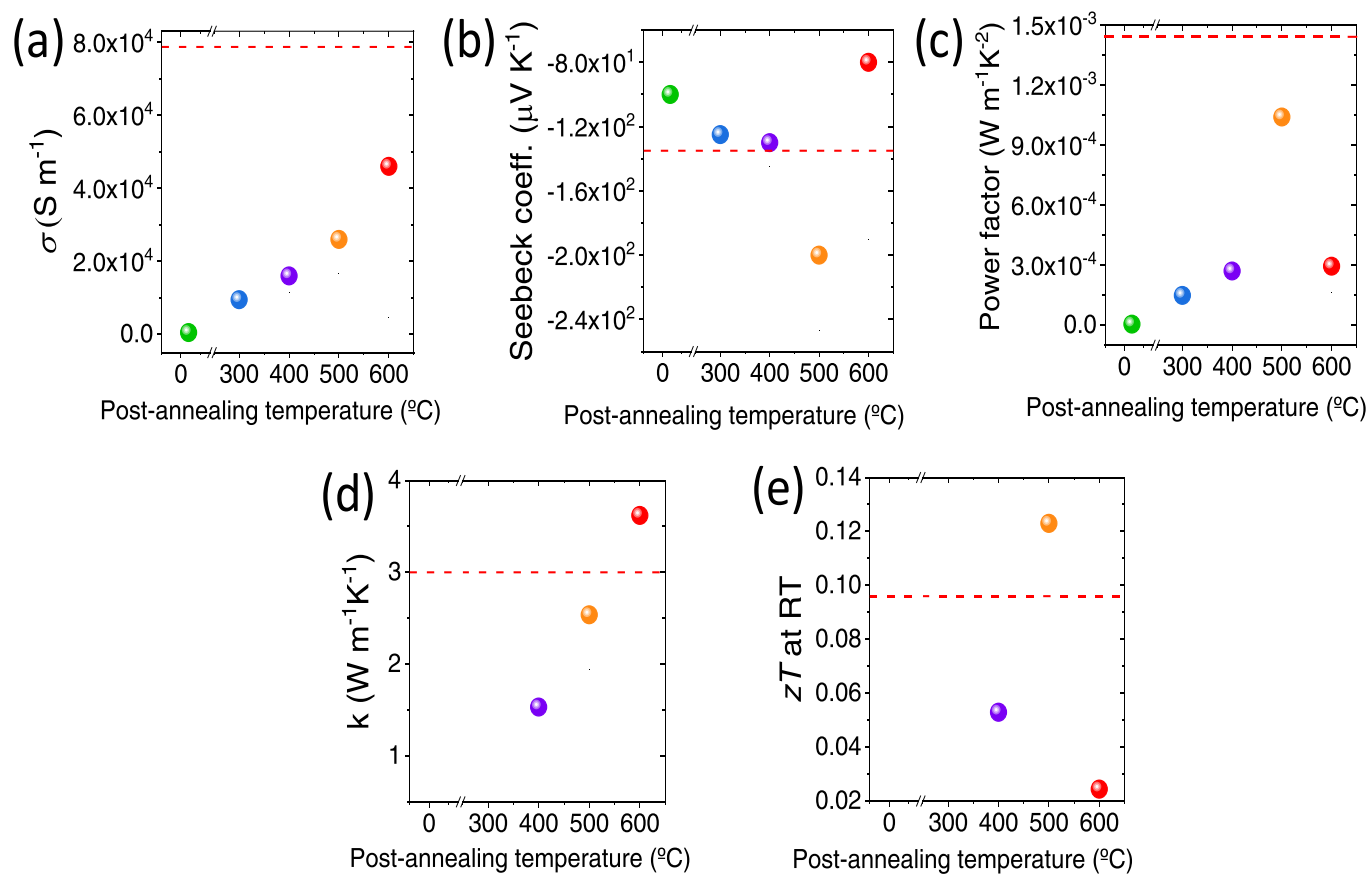

**Fig. 2.** Measured values for the (a) electrical conductivity (associated error 5 %), (b) Seebeck coefficient (associated error 10 %), (c) power factor, and (d) thermal conductivity (associated error 15 %) of the different CSP sintered samples as a function of the annealing temperature of the ulterior thermal treatment. In all the graphs, the dashed red line represents the value obtained for SPS sintered samples from the same starting powders. (e) Calculated figure of merit, again as a function of the post-annealing temperature and showing the published calculated value for an SPS sintered sample. The SPS was conducted at 600 °C with a dwell time of 15 min, heating rates of 120 °C min$^{-1}$ by using 50 MPa of pressure in Ar atmosphere (from Ref. [6]).

Structural parameters such as the lattice dimensions and the volume-averaged crystallite size for the different identified phases have been determined. Specifically, the crystallite size for $CoSb_3$ (see Fig. 4b) shows a similar trend than that obtained for the particle size calculated from FESEM measurements, with an exponential increase as the post-treatment temperature rises. The starting powder exhibits a crystallite size of around 23.9(2) nm, reaching values of about 212(4) nm at the highest post-treatment temperature of 600 °C (one order of magnitude more). At this point, it should be remarked that the particle size obtained from FESEM (Fig. 1g) and the crystallite size for $CoSb_3$ determined from Rietveld refinements (Fig. 4b) show the same values for the starting powder and the sample NC, suggesting the monodomain structure of the $CoSb_3$ grains. At the lowest post-annealing temperatures of the sample NC, the particle and the crystallite size of $CoSb_3$ phase present small discrepancies which increase significantly at 600 °C, where the particle size is three times bigger than the crystallite size. These differences at higher post-annealing temperatures could be due to the compositional variations or to the fact that the different crystalline directions do not have time to rearrange themselves during the post-annealing process as consequence of the grain growth. The possible presence of defects due to crystal orientation mismatches could be responsible for leaving a reduction of the thermal conductivity by phonon scattering (see Fig. 2) and explain that the behavior is linear and not exponential as the grain growth does [47,50].

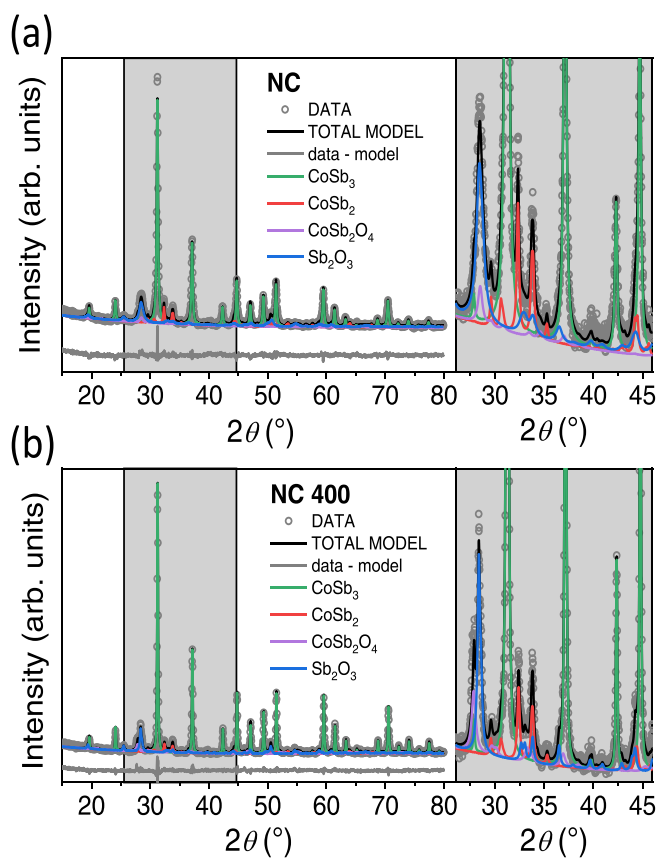

**Fig. 3.** PXRD data collected experimentally along with corresponding Rietveld models for samples (a) NC and (b) NC 400. Besides the total Rietveld model (black) and models for the individual phases are also shown, i.e., $CoSb_3$ (green), $CoSb_2$ (red), $CoSb_2O_4$ (purple) and $Sb_2O_3$ (blue). On the right hand side, the lower intensity region in the 2θ range 26–46° has been magnified to better illustrate the contribution from the individual phases to the total diffraction pattern.

From a structural point of view, important modifications on both the Co, Sb and Te coordination and the short-range order could have happened during the sintering process of the samples, which would affect their functional response. To elucidate the influence of the CSP and the post-annealing process, both XANES and EXAFS investigations were performed on densified nanocomposites and compared with the starting powders.

XANES measurements were carried out at the Co, Sb and Te K-edge of the nanocomposites densified by CSP, along with the starting powders and metallic foils used as reference for comparison, which are shown in Figs. 5a, 5c, and 5e, respectively. At the Co K-edge, the absorption signal of the starting powders and samples can be associated mainly with $CoSb_3$, where the absorption edge in the 7707–7714 eV range and the main XANES features can be identified in all cases with similar characteristics [51]. The first inflection point of the main Co K-absorption edge is associated with 1s-4p transitions [52,53] and its position is represented in Fig. 5b along with the position of the absorption edges for metallic Co foil, CoO and $Co_3O_4$ references. Comparing the position of the main absorption edge of experimental data for samples and standard compounds, the average oxidation state of the Co absorbing atom for all samples is estimated. A similar position of the absorption edge is observed for the starting powder and all samples, regardless of the annealing temperature. In all cases, values fall in energies near the metallic Co foil (7709 eV) [54], indicating a similar average valence for all samples, close to $Co^0$. The slight differences at the edge position and the characteristics of the XANES signal with respect to the metallic Co reference may be associated with different coordination or oxidation of the Co absorbing atoms [55]. The average of oxidized Co atoms is minimal considering the $CoSb_2O_4$ fraction estimated by PXRD technique. In addition, the intensity of the absorption edge is lower in comparison with that of the Co foil and is slightly enhanced as the CSP is carried out from the starting powders and after the post-annealing temperature increases, as displays the inset in Fig. 5a, which is an indication of the structural disorder induced by the sintering process. With respect to the relative intensity of the main resonances after the absorption edge position, these change a bit with the CSP and the post-annealing temperature. The first oscillation after the absorption edge around 7718 eV increases from the starting powder to the CSP and the post-annealings, reaching the maximum at 600 °C. Similar behavior is identified in other absorption resonances, except for the second and the third ones at 7724 and 7730 eV, respectively, in which the CSP induces an intensity increase and the post-annealings decrease the intensity. These findings suggest modifications in the coordination of the Co atoms related to the presence of a different local environment around Co atoms and associated with the compositional changes after the CSP and the post-thermal treatment process. The most significant changes are identified for the sample NC 600 where $CoSb_3$ decreases to 67 wt% and $CoSb_2$ increases to 30 wt%, determined by PXRD experiments.

The Sb K-edge XANES spectra are displayed in Fig. 5c. The absorption edge and main peak can be attributed to 1s →5p transitions, while the absorption peaks at greater energy side correspond to the transitions from the Sb 1s core-level states to the unoccupied Co 4p states due to the existence of $d^2sp^3$ orbital hybridization in $CoSb_6$ octahedron [56]. The position of the first inflection point falls at similar positions for all nanocomposites and that of the starting powders, at higher energies than the metallic Sb foil, as Fig. 5d shows. This shift towards larger energies may be related to a fraction of Sb absorbing atoms with different coordination or oxidation state [57], although the values are within the error. Considering the PXRD results, a percentage of about 14–21 wt% of oxidized Sb atoms as crystalline $CoSb_2O_4$ and $Sb_2O_3$ phases is determined (see Fig. 4a). The lowest fraction of oxidized Sb atoms corresponds to the starting powders, whose composition could contain some amorphous phase generated probably by the mechanosynthesis process and crystallized during the CSP. Specifically, the sample post-annealed at 600 °C (NC 600) does not present any oxidized phase by PXRD and the presence of amorphous phases at these temperatures is rather unlikely. Thus, the energy shift in Fig. 5d could be attributed to different coordination of the Sb absorbing atoms. In addition, except for NC 600, no significant changes are noted in the absorption edge intensity (at 30491 eV), which would indicate no variations in the unoccupied states [58,59]. Specifically, as the post-thermal treatment of the CSP nanocomposite is performed at 600 °C, the whiteline of the XANES spectrum at 30499 eV shows an intensity decrease and new resonances at higher energies, indicating different coordination in the nanocomposite generated likely by the presence of a larger fraction of $CoSb_2$ phase (30 wt% determined by PXRD experiments).

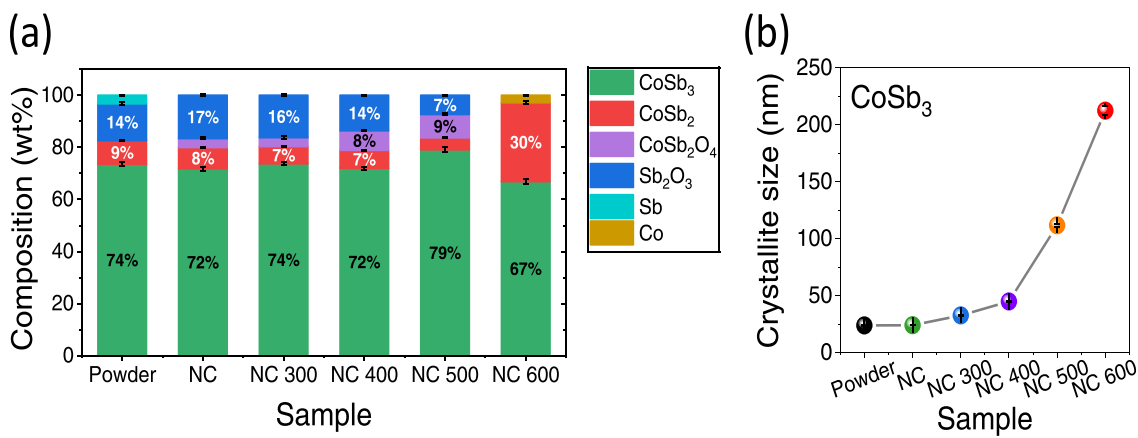

**Fig. 4.** Refined values extracted from Rietveld analysis: (a) weight fractions for the individual phases, following the color code: $CoSb_3$ (green), $CoSb_2$ (red), $CoSb_2O_4$ (purple), $Sb_2O_3$ (blue), Sb (light blue) and fcc-Co (golden). In this figure, only weight fractions > 5 % are labelled. (b) Volume-weighted average crystallite sizes for the majority phase, $CoSb_3$.

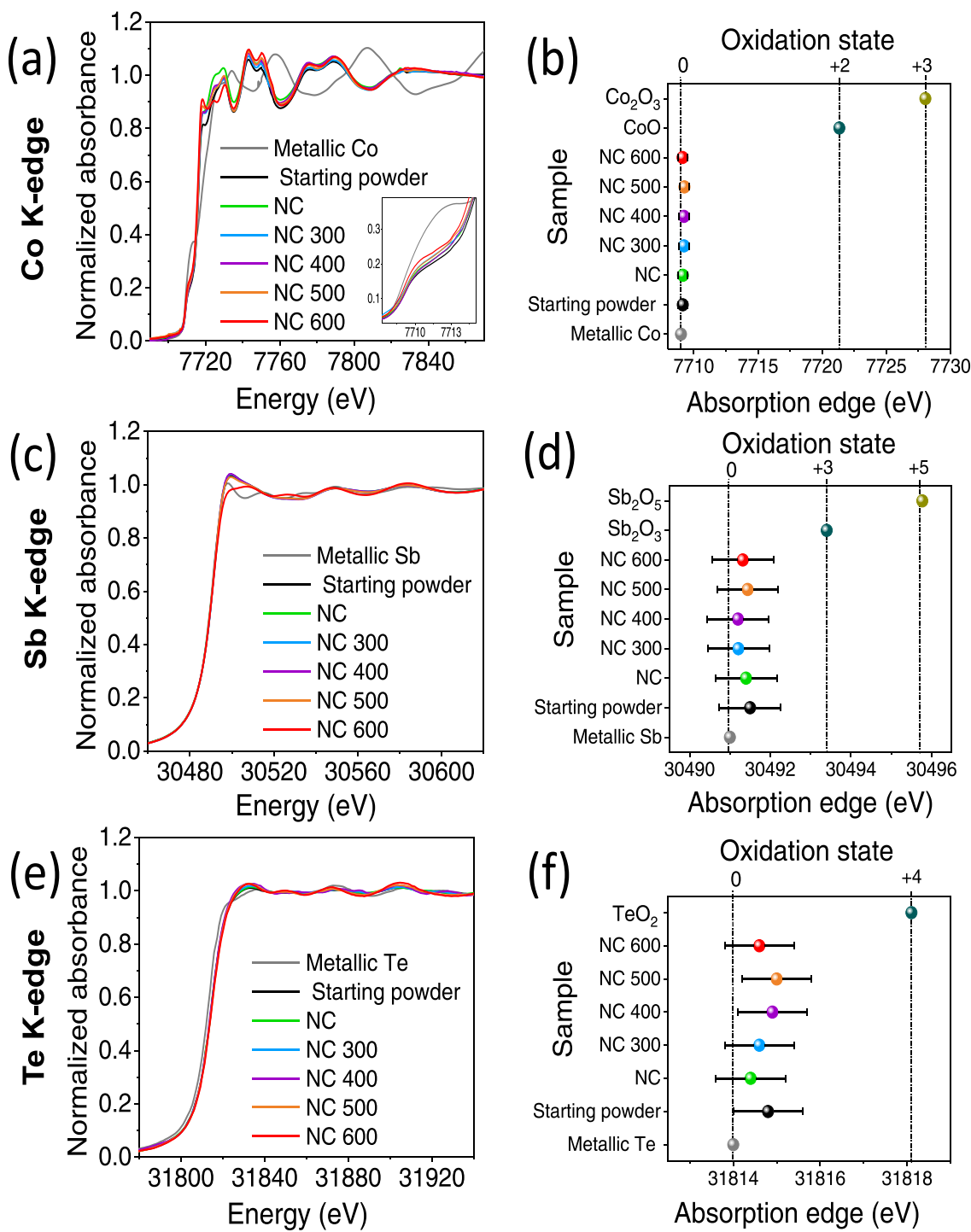

**Fig. 5.** XANES spectra at (a) Co K-edge, (c) Sb K-edge and (e) Te K-edge. Average valence from the position at the absorption edge in XANES region at (b) Co K-edge, (d) Sb K-edge and (f) Te K-edge. The average oxidation state of different absorbing atoms is indicated for metallic references (valence 0) and the oxide powder references: CoO, $Co_2O_3$, $Sb_2O_3$, $Sb_3O_5$ and $TeO_2$.

Fig. 5e shows the XANES spectra at the Te K-edge for all nanocomposites along with the metallic Te reference. In all samples and the starting powders, the energy of the main absorption and the absorption whitelines are similar with several differences with respect to the Te reference. The position of the absorption edge is represented in Fig. 5f and it is similar for all nanocomposites and to that of the starting powders. A slight shift at higher energies with respect to that of the Te metal (see Fig. 5e and f) can be related to the different coordination of the Te absorbing atoms with respect to the metallic Te. The presence of any Te oxidized compound, not determined by PXRD, cannot be discarded. For all nanocomposites, the absorbing characteristic indicates that the unoccupied 5p states are located close to the Fermi level. A possible 5p-orbital hybridization between Te and Sb may occur in the nanocomposites, which results in a charge transfer from Sb to Te and the enhancement of p-d orbital hybridization between Co and Sb, as previously reported for In-doped $CoSb_3$ systems [60]. As the post-annealing increases a trend is identified: larger absorption intensity at the whiteline and more defined resonances suggesting greater coordination and a more crystalline nanocomposite with the post-annealing temperature.

The short-range order of cations around the Co, Sb and Te ions and the neighbour bondlengths are analyzed as a function of the influence of the sintering process with the CSP and the post-annealing process in each case using EXAFS technique to complement the PXRD and XANES results.

Fig. 6 displays the modulus of the Fourier transform (FT) of the EXAFS signal at the Co, Sb and Te K-edge for samples prepared by CSP and those post-annealed, along with the starting powders and references. The FT is performed in the $k^3\chi(k)$ weighted EXAFS signal between 2.6 and 12.4 Å$^{-1}$, 2.6 and 13.0 Å$^{-1}$, and 2.6 and 10.0 Å$^{-1}$ at the Co, Sb, and Te K-edge, respectively. Experimental EXAFS results are fitted in $R$-space in the range 1.5–3.0 Å at the Co and Te K-edges and 1.2–3.0 Å at the Sb K-edge using the FEFFIT code [61]. The fitting was performed by fixing the shift at the edge energy $E_0$ for each absorption edge, which was previously calculated from the starting powder used as reference. Hence, the amplitude $A$, the interatomic distance $R$ and the Debye-Waller (DW) factors $\sigma^2$ are used as free parameters. For the fitting, a shell produced by the interaction of a Co absorbing atom with Sb atoms, two shells from an Sb absorbing atom with O and Co atoms, and a shell from the interaction of a Te absorbing atom surrounded by Co atoms are considered at the Co, Sb and Te K-edge, respectively. Taking into account that several compounds have been identified by PXRD in the nanocomposites, the interpretation of coordination number $N$ is not easy and is done in a semiquantitative way. Table 2 displays the numerical EXAFS results obtained from the fitting for Te-doped $CoSb_3$ nanocomposites prepared by CSP, post-annealed as well as the results for the starting powders.

At the Co K-edge, the EXAFS signal of samples presents one shell around 2.52 Å which is attributed to Co-Sb neighbour distances related to the majority $CoSb_3$ compound, which appears at larger distances than the Co metallic reference (see Fig. 6a). Comparing the results between the different nanocomposites, similar positions and DW factors are identified for the Co-Sb shell. A slight increase of the amplitude is noted from 2.88 for the starting powder to 3.72 for the nanocomposite post-annealed at 600 °C (NC 600), as Table 2 shows, which can relate to the increase in the coordination and/or the crystallization of compounds based on Co.

Fig. 6b displays the EXAFS spectra at the Sb K-edge. An Sb-O shell around 1.99 Å and an Sb-Co shell at 2.52 Å are identified in all nanocomposites. The first maximum could be related to oxidized absorbing atoms in compounds identified in the nanocomposites by PXRD, such as the $Sb_2O_3$ and $CoSb_2O_4$ (see Figs. 3 and 4). However, due to the low intensity of this Sb-O shell, the proportion has to be small in accordance with the PXRD and XANES results. For the coordination of the Sb-O shell, an increase is noted up to the post-annealing temperature of 400 °C from which a decrease is identified, leaving a shell with very low intensity, which could indicate the removal of oxidized materials as obtained by PXRD. For the Sb-Co distance, the coordination increases progressively up to the composite post-annealed at 600 °C with the highest amplitude. This latter amplitude increase may be showing a growth in the coordination number or further crystallization of the structure supporting the PXRD data.

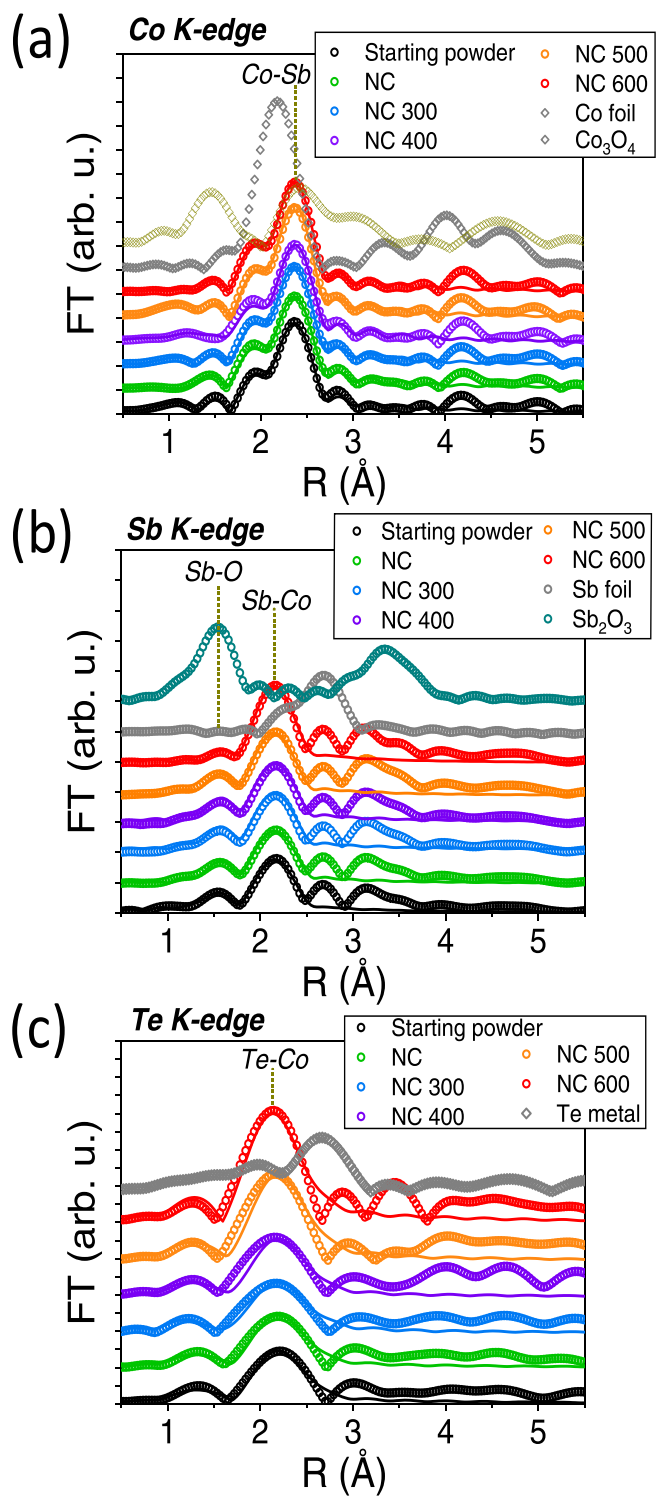

**Fig. 6.** Fourier transform modulus of the EXAFS spectra at the (a) Fe and (b) Sb and (c) Te K-edge and best-fitting simulations (continuous lines) of the $CoSb_3$-based composited processed after the CSP and that post-annealed at 300, 400, 500 and 600 °C. EXAFS spectra of the starting powders, along Co, $Sb_2O_3$, Sb and Te references are included.

**Table 2**
Results of the EXAFS fittings for the first and second shells of thermoelectric nanocomposites prepared by CSP and compared with starting powder at Co, Sb and Te K-absorption edges.

| Sample | Absorption edge | Shell | N·A | R (Å) | DW (Å$^2$) |
|---|---|---|---|---|---|
| **Starting powder** | **Co K-edge** | **Co-Sb** | 2.9(5) | 2.524(6) | 0.0033(9) |
|  | **Sb K-edge** | **Sb-O** | 0.77(9) | 1.98(4) | 0.004(1) |
|  |  | **Sb-Co** | 1.2(2) | 2.520(8) | 0.003(1) |
|  | **Te K-edge** | **Te-Co** | 2.1(5) | 2.80(1) | 0.006(1) |
| **NC** | **Co K-ege** | **Co-Sb** | 3.1(3) | 2.522(2) | 0.0036(5) |
|  | **Sb K-edge** | **Sb-O** | 0.81(7) | 1.99(2) | 0.004(1) |
|  |  | **Sb-Co** | 1.2(2) | 2.521(7) | 0.003(1) |
|  | **Te K-edge** | **Te-Co** | 2.2(4) | 2.80(1) | 0.006(1) |
| **NC 300** | **Co K-ege** | **Co-Sb** | 3.2(1) | 2.520(2) | 0.0034(6) |
|  | **Sb K-edge** | **Sb-O** | 0.84(9) | 1.99(3) | 0.004(1) |
|  |  | **Sb-Co** | 0.6(1) | 2.520(6) | 0.003(1) |
|  | **Te K-edge** | **Te-Co** | 2.3(4) | 2.80(1) | 0.006(2) |
| **NC 400** | **Co K-ege** | **Co-Sb** | 3.5(5) | 2.520(3) | 0.0039(1) |
|  | **Sb K-edge** | **Sb-O** | 0.89(9) | 1.99(3) | 0.004(1) |
|  |  | **Sb-Co** | 1.3(2) | 2.520(6) | 0.003(1) |
|  | **Te K-edge** | **Te-Co** | 2.5(4) | 2.80(1) | 0.006(1) |
| **NC 500** | **Co K-ege** | **Co-Sb** | 3.7(1) | 2.522(2) | 0.0036(6) |
|  | **Sb K-edge** | **Sb-O** | 0.76(8) | 1.99(3) | 0.004(1) |
|  |  | **Sb-Co** | 1.4(2) | 2.518(6) | 0.003(1) |
|  | **Te K-edge** | **Te-Co** | 3.8(5) | 2.80(1) | 0.006(1) |
| **NC 600** | **Co K-ege** | **Co-Sb** | 3.7(2) | 2.521(1) | 0.0037(3) |
|  | **Sb K-edge** | **Sb-O** | 0.26(8) | 2.00(9) | 0.004(1) |
|  |  | **Sb-Co** | 1.7(2) | 2.521(6) | 0.003(1) |
|  | **Te K-edge** | **Te-Co** | 4.7(5) | 2.79(2) | 0.006(1) |

EXAFS spectra at the Te K-edge are presented in Fig. 6c along with that of the Te metallic reference. Here the main peak is obtained at 2.80 Å, at a lower position than the Te-Te metallic bonding of the metallic Te reference [62], and it is related to a Te-Co shell considering that the Te absorbing atoms are occupying the Sb sites. If we consider a Te-Sb shell with the Te atoms occupying the Co sites, the distances would appear at higher values and a poor fitting is reached. Likely, the occupation of Te atoms in Sb sites may be related to the similar atomic radius they possess with respect to Co, which has a smaller radius. With the post-annealing process, a higher shell amplitude with the temperature is obtained that may be attributed to a better redistribution of the Te atoms with the annealing and/or a better crystallization of materials, as was noted above. Regarding the neighbor distances and the DW factor, no differences are observed during the sintering process at short-range ordering of Te ions.

The combined X-ray absorption study at the Co, Sb and Te K-edges shows the presence of the CoSb$_3$-structure in all cases, with a lower fraction for the sample NC 600. The average valence was mainly determined as 0 for the three absorbing atoms, with a small fraction of oxidized compounds at the Sb K-edge as obtained by PXRD. The largest structural variation is obtained for the sample post-annealed at 600 °C. In addition, the investigations confirm that Te ions replace Sb in the cubic frame of the skutterudite structure to form a CoSb$_{2.75}$Te$_{0.25}$ phase.

These compositional and structural modifications, along with the changes in the morphological characteristics, are dependent on the sintering conditions and explain the thermoelectric properties of nanocomposites. On one hand, the grain size affects both the electrical and thermal conductivity, which increase as the grain size increases. When the grains size increases, the grain boundaries density reduces, and thus, these scatter the phonons less and the nanostructuration is not so effective for reducing the conductivities. In addition, the discrepancy between the crystallite size and the grain size can lead to defects between the crystals reducing the thermal conductivity, as commented above, yielding a less pronounced growth than the grain size with treatment temperature. On the other hand, the fraction of CoSb$_3$ present in the different nanocomposites is reflected in the Seebeck coefficient. The highest value is measured for sample NC 500, which presents the highest contents of the CoSb$_3$ phase, while the decrease in the CoSb$_3$ content of the CSP nanocomposite post-annealed at 600 °C and the increase in the CoSb$_2$ phase leaves an abrupt decrease of the Seebeck coefficient. Maximum values of power factor and figure of merit are obtained for the nanocomposite post-annealed at 500 °C, where a compromise is found between the increase of the grain size with annealing temperature (which increases the electrical conductivity) and the content of the CoSb$_3$ phase (reaching the maximum content and, thus, the maximum Seebeck coefficient).

Therefore, the CSP followed by a post-annealing step at 500 °C in Ar atmosphere has allowed achieving dense CoSb$_3$-based composites with a relative density of 92 % and improved thermoelectric properties (zT of 0.12(3) at RT) close to the operating temperatures. The sintering approach using a non-aqueous solvent in the process retains the oxidation of the initial powders, leaving a controlled grain growth and crystallite size as well as the largest fraction of CoSb$_3$. These results show a cost-effective and easily scalable sintering route and open the way to study thermoelectric materials under working conditions and direct conformation.

## 4. Conclusions

CoSb$_3$-based thermoelectric nanocomposites have been sintered by CSP with suitable mechanical integrity and similar morphological and structural properties than the starting powders for the first time. A post-annealing process in Ar atmosphere enhances the relative density of samples to values around 92 % at temperatures ≥ 500 °C, identifying a grain and crystallite size growth as well as the partial compositional transformation with an increase of the secondary phase CoSb$_2$ phase up to 30 wt% at 600 °C. The thermoelectrical properties are dependent on the sintering process, with the highest power factor (1000(200) μW m$^{-1}$ K$^{-2}$) and zT (0.12(3) at RT) for the nanocomposites sintered by CSP plus a subsequent post-annealing at 500 °C, which presents the highest CoSb$_3$ phase content with adequate grain and crystallite size. For the highest post-annealing temperature studied, 600 °C, the content on CoSb$_3$ is reduced, favouring the appearance of CoSb$_2$ and cobalt phases, reducing the effective Seebeck coefficient and thus, the thermoelectric efficiency. In all cases, the average oxidation state of Co, Sb and Te absorbing atoms basically remains during the sintering process, determining some variations in short-range order associated with an increase in the coordination number of the absorbing atoms or the crystallization of the structure. These outcomes imply the incorporation of a new methodology for the cost-effective fabrication of nanostructured thermoelectric skutterudite-based materials at reduced process temperatures with a great potential for numerous applications.

**Author Contributions**

All authors have approved the final version of the manuscript.

**Data Availability**

Data will be made available on request.

**Declaration of Competing Interest**

The authors declare that they have no known competing financial interests or personal relationships that could have appeared to influence the work reported in this paper.

**Acknowledgements**

This work has been supported by the Ministerio Español de Ciencia e Innovación (MICINN) through the projects MAT2017-86540-C4-1-R, MAT2017-86540-C4-3-R, RTI2018-095303-A-C52, PID2020-114192RB-C41 and PID2020-118430GB-100. Part of these experiments were performed at the CLAESS-BL22 beamline at ALBA Synchrotron (Grant no. 2020094658) and in collaboration with ALBA staff. A.S. and C.V.M. acknowledge financial support from the Comunidad de Madrid for an "Atracción de Talento Investigador" Contract no. 2017-t2/IND5395 and 2019-t1/IND-13541, respectively. C.G.-M. acknowledges financial support from MICINN through the "Juan de la Cierva" program (FJC2018-035532-I). A.M. is indebted to MICINN for the "Ramón y Cajal" contract (RYC-2013-14436), which is co-financed with European Social Fund.

**Appendix A. Supporting information**

Supplementary data associated with this article can be found in the online version at doi:10.1016/j.jallcom.2022.167534.